\documentclass[]{beilstein}
\usepackage{graphicx,amsmath,color}
\usepackage{amssymb}
\usepackage{sidecap}
\usepackage{cleveref}
\usepackage{hyperref}
\usepackage{epstopdf}
\newcommand{\wc}{W(CO)$_6$}
\newcommand{\fc}{Fe(CO)$_5$}

\newcommand{\coco}{Co$_2$(CO)$_8$}

\newcommand{\cofour}{Co(CO)$_4$}

\newcommand{\cotwelve}{Co$_4$(CO)$_{12}$}

\newcommand{\bc}{${\beta}${-}cristobalite}
\newcommand{\sio}{SiO$_2$}
\newcommand{\FOH}{FOH - SiO$_2$}
\newcommand{\POH}{POH - SiO$_2$}
\newcommand{\pt}{CH$_3$C$_5$H$_5$Pt(CH$_3$)$_3$}
\begin{document}
\title{Spontaneous Dissociation of {\coco} and Autocatalytic growth of Co on {\sio} : A Combined Experimental and Theoretical Investigation}
\author{K. Muthukumar}
\affiliation{Institut f\"ur Theoretische Physik, Goethe-Universit\"at, Max-von-Laue-Stra{\ss}e 1, 60438 Frankfurt am Main, Germany}
\author[1]{H. O. Jeschke}
\author*[1]{R. Valent\'\i}{valenti@itp.uni-frankfurt.de}
\author{E. Begun}
\affiliation{Physikalisches Institut, Goethe-Universit\"at, Max-von-Laue-Stra{\ss}e 1, 60438 Frankfurt am Main, Germany}
\author[2]{J. Schwenk}
\affiliation{Present Address: Empa, CH-8600 D\"{u}bendorf, Switzerland}
\author[2]{F. Porrati}
\author*[2]{M. Huth}{michael.huth@physik.uni-frankfurt.de}
\maketitle
\begin{abstract}
  We present experimental results and theoretical simulations of the
  adsorption behavior of the metal-organic precursor {\coco} on {\sio}
  surfaces after application of two different pre-treatment steps,
  namely by air plasma cleaning or a focused electron beam
  pre-irradiation. We observe a spontaneous dissociation of the
  precursor molecules as well as auto-deposition of cobalt on the
  pre-treated {\sio} surfaces. We also find that the differences in
  metal content and relative stability of these deposits depend on the
  pre-treatment conditions of the substrate. Transport measurements of
  these deposits are also presented. We are led to assume that the
  degree of passivation of the {\sio} surface by hydroxyl groups is an
  important controlling factor in the dissociation process. Our
  calculations of various slab settings using dispersion corrected
  density functional theory support this assumption. We observe
  physisorption of the precursor molecule on a fully hydroxylated
  {\sio} surface (untreated surface) and chemisorption on a partially
  hydroxylated {\sio} surface (pre-treated surface) with a spontaneous
  dissociation of the precursor molecule.  In view of these
  calculations, we discuss the origin of this dissociation and the
  subsequent autocatalysis.
\end{abstract}
\section{Introduction}
In recent years, focused electron beam induced deposition (FEBID) has
emerged as a versatile, high-resolution technique for nanostructure
fabrication in contrast to the more conventional nanolithographic
techniques. In FEBID a previously adsorbed precursor gas is
dissociated in the focus of an electron beam. The non-volatile part of
the dissociation products remains as a deposit whose shape and
position can be accurately controlled by the lateral positioning of
the electron beam in an electron microscope
\cite{utke:1197,Wnuk2011257,ANIE:ANIE201002677,doi:10.1080/10408430600930438,ebidsilvis}.  Mostly gaseous ({\wc}, {\fc}, and
{\pt})\cite{0957-4484-20-19-195301,porrati:063715,1367-2630-11-3-033032,Shen2012}
but also liquid organometallic precursors (chloroplatinic acid)
\cite{doi:10.1021/nl9012216} are being used to deposit metals or metal
composites on selected regions of the substrates.  Deposits with a
wide spectrum of properties and composition can be consequently
obtained due to the availability of suitable
precursors \cite{Wnuk2011257,utke:1197}.

{\coco} has been recently used as a precursor molecule in FEBID to
obtain granular deposits with differing compositions of
cobalt \cite{0957-4484-20-37-372001}. Electronic and physical
properties, such as grain size and metal content of these deposits,
depend strongly on the deposition and pre-treatment conditions of the
substrate. By regulating these conditions, deposits of desired size
and different Co content can be
fabricated \cite{lau:1295,0957-4484-20-47-475704,ANIE:ANIE201004220,Utke2004553}.
For example, granular Co-nanostructures suitable for micro Hall
sensing devices\cite{0957-4484-21-11-115503} were thus obtained. Very
recently this precursor has also been used in combination with the
precursor {\pt} to fabricate nanogranular CoPt-C structures with CoPt
nanocrystallites having the L1$_0$ crystal structure with
hard-magnetic properties. \cite{0957-4484-23-18-185702}  Also, it has
been shown that under well-controlled conditions, Co line structures
with a width down to 30 nm are
feasible \cite{doi:10.1021/nn201517r,nikulina:142401}. These findings
make FEBID with the Co-precursor particularly attractive for the
fabrication of micromagnetic structures in the sub-100 nm regime,
relevant for studies of the domain wall dynamics
\cite{fernandez-pacheco:192509}, the Barkhausen effect in single
domain wall structures \cite{das:042507} and dipolar coupling effects.
\cite{porrati:3157}

While several experimental studies based on Infra-red spectroscopy 
~\cite{Suvanto199925,doi:10.1021/ic00194a038,kmrao1988,Rao1988466} 
and theoretical~\cite{EJIC:EJIC3031,Suvanto2000277,Barckholtz2000212,doi:10.1021/jp9728565}
studies on {\coco} are available in the literature
an issue that remains unclear so far is the possible tendency of this
precursor to spontaneously dissociate on {\sio} surfaces, as well as
to autocatalytically grow by spontaneous decomposition on existing Co
clusters. Similar features have been reported to be exhibited by
{\fc} \cite{ANGE:ANGE201001308,hochleitner:939}. In order to evaluate the previous
effects in the FEBID process, it is mandatory to acquire an in-depth
knowledge of the interactions between the precursor molecule {\coco}
and {\sio} surfaces representing the different pre-treatment
conditions of the substrate.~\cite{PhysRevB.84.205442} In the present
work, we report on experimental results of Co deposition by
spontaneous dissociation of the precursor {\coco} on untreated and two
differently pre-treated {\sio} surfaces (by an air plasma cleaning
process and a pre-growth electron irradiation of selected areas).
To our knowledge, no systematic theoretical studies 
with in-depth DFT calculations on {\coco} adsorbed on
different {\sio} surfaces are available. Therefore, we extent the study using density functional theory (DFT)
calculations on slabs representing the various {\sio} surface
conditions and aim to relate the observations to the plasma and electron 
irradiation conditions prevailing in FEBID experiments. 

\section{Experimental Details}\label{Experiment}
Cobalt growth and imaging experiments were carried out at room
temperature in a Dual Beam scanning electron microscope (FEI Nova
NanoLab 600) with a Schottky electron emitter. A plasma source using
ambient air at a chamber pressure of $1\times 10^{-4}$ to $5\times
10^{-4}$ mbar was used for the surface activation experiment (GV10x
Downstream Asher, ibss Group). Electron pre-growth irradiation
experiments were carried out at 5 kV beam voltage and 1.6 nA beam
current. Si (100) (p-doped) substrates with thermal oxide layers of 50
nm up to 285 nm were used. Before use, the substrates were chemically
cleaned by acetone, isopropanol and distilled water in an ultrasound
bath.

In the plasma activation experiments the silica sample surface (285 nm
oxide layer) was exposed to the plasma discharge for 75 min after the
scanning electron microscope (SEM) chamber had been evacuated to its
base pressure of about $5\times 10^{-6}$ mbar. After the plasma
treatment the chamber was again evacuated to base pressure and
Co-precursor flux was admitted to the chamber by opening the valve of
a home-made gas injection system for 30 min causing a pressure
increase to $3\times 10^{-5}$ mbar which dropped within ten minutes to
about $6\times 10^{-6}$ mbar.  The gas injection system employs a
stainless steel precursor capsule with a fine-dosage valve.  The
precursor temperature was set by the ambient conditions to 27$^{\circ}$C.
From the known precursor temperature and associated vapor pressure, as well as the geometry of 
our gas injection system we can roughly estimate the maximum molecular flux at the substrate 
surface to be $1.4\times 10^{17}$ cm$^{-2}$s$^{-1}$ following Ref{\cite{Friedli2009}}.

In the second series of experiments the untreated silica surface was
pre-growth irradiated with a focused electron beam which was moved in
a raster fashion (dwell time 100 $\mu$s, pitch 20 nm) for 30 min 
over a rectangular region of 3.7 x 1.0 $\mu$m bridging the gap between two 
pre-patterned Cr/Au electrodes. The background pressure during the 
irradiation process was $6\times 10^{-6}$ mbar. 
Within the 30 min irradiation time about 2000 passes of the 
rectangular pattern were performed amounting to an overall dose of 0.78 $\mu$C/$\mu$m$^2$. 
After this treatment the Co-precursor
was admitted to the SEM chamber and the current between the electrodes
was measured at a fixed bias voltage of 10 mV as a function of time
(see Fig.~\ref{SEM}~(b)).  By this method the formation of a
conducting path between the metallic electrode can be conveniently
followed and gives a first indication of the spontaneous formation of
a deposit.  After about 20 min the injection was stopped, the SEM
chamber was flushed with dry nitrogen and evacuated again for image
acquisition.

\section{Computational Details}\label{Formalism}

We performed spin polarized density functional theory (DFT)
calculations within the generalized gradient approximation in the
parametrization of Perdew, Burke and Ernzerhof
(PBE).~\cite{PhysRevLett.77.3865, PhysRevLett.78.1396} Corrections for
long range van der Waals interactions \cite{JCC20495, wu8748} were
included in all calculations. We worked with a kinetic energy cut-off of
400~eV and relaxed all the ions with the conjugate gradient scheme until
the forces were less than 0.01~eV/{\AA}. In order to reproduce the
experimental settings, untreated {\sio} surfaces were described in
terms of fully hydroxylated substrates, while pre-treated {\sio}
surfaces were described in terms of partially hydroxylated substrates \cite{Mens1996133,Idriss,Gopel1984333}.
Our (fully and partially hydroxylated) {\sio} substrates consist of
four layers of ($3\times 3$) supercells of {\bc} primitive unit cells.
We calculated total energy differences ${\Delta}E$ for substrates,
precursor molecules and the complex of substrate with adsorbed
precursor molecules as reported previously~\cite{PhysRevB.84.205442,Shen2012}
using the projector augmented wave method~\cite{PhysRevB.59.1758, PhysRevB.50.17953} as
implemented in the Vienna Ab-initio Simulation Package
(VASP) \cite{Kresse1996, Kresse1993, Kresse1994}. In the geometry
optimizations for the molecule and the substrate models the Brillouin
zone was sampled at the ${\Gamma}$ point only. In addition, to analyse
the molecular orbitals, we employed Turbomole 6.0~\cite{treutler:346,
Eichkorn1997} to optimize the {\coco} molecule with triple-zeta
valence plus polarization basis sets with the PBE
functional using the resolution-of-the-identity (RI) approximation.
The Bader charge partition analysis was performed
using the code of Henkelman et al. to determine the charges
of individual atoms \cite{bader,Henkelman2006354}.

\section{Results and Discussion}
\subsection{Formation of Co from {\coco} on pre-treated {\sio} surfaces}
\begin{figure}
\begin{center}
\sglcolfigure{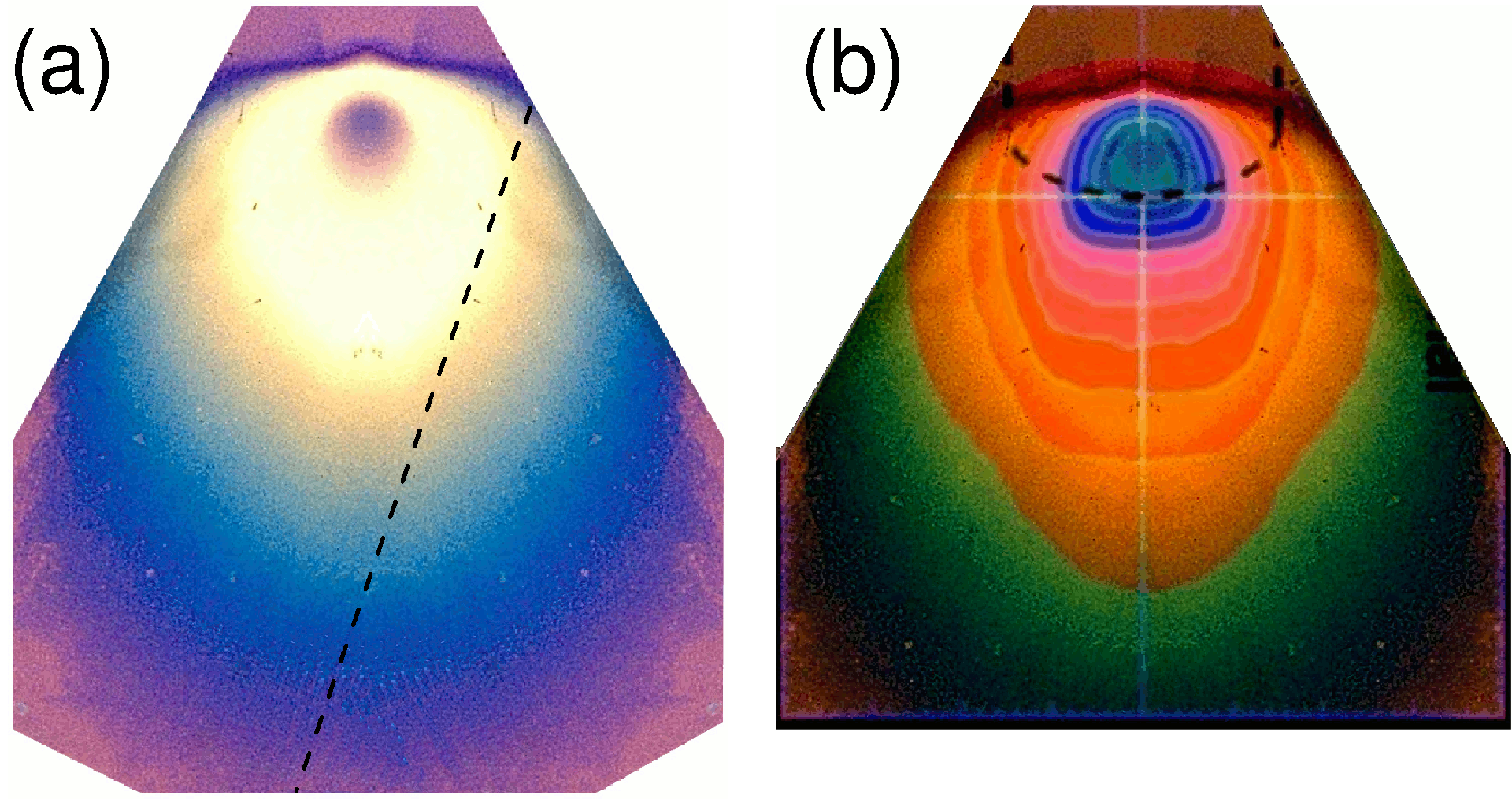}
\caption{(Color online) (a) Optical micrograph of the Co dissociation
product on the plasma-activated silica surface. The deposit mimics the flux profile set by the gas injection
needle. The dashed line represents the rightmost substrate edge. The deposit profile to the right of the dashed
line was complemented by image processing from the left side for ease
 of comparison.
(b) Overlay of the calculated precursor flux profile from Ref.~\cite{Friedli2009}
(contour lines) with the isotropically scaled
optical microsocope image of the deposit profile shown in (a).
}\label{optics}
\end{center}
\end{figure}

In Fig.~\ref{optics}~(a) we present an optical micrograph of a
spontaneous dissociation product obtained on the plasma pre-treated
{\sio} surface. A Co-rich layer of varying thickness has formed whose
lateral shape clearly depicts the precursor flux profile imposed by
the gas injection needle.  This profile appears in
Fig.~\ref{optics}~(b) and is in excellent correspondence with
simulations of the precursor flux presented in
Ref. ~\cite{Friedli2009}. It should be stressed that no such
spontaneous growth was observed on the untreated {\sio} surface after
30 min exposure to the Co-precursor.  
Only for extended exposure times (30 min and more) we find also on 
the untreated surfaces evidence for the
tendency of spontaneous dissociation. At this stage we are led to
assume that the untreated {\sio} surface, usually hydroxylated after
chemical cleaning as performed by us, shows a weak tendency to induce
spontaneous dissociation of the Co-precursor. Partial or full removal
of the hydroxyl surface passivation layer leads to an increased
driving force for dissociation. This will be discussed in more detail
in the the next section where we present results obtained in the
framework of DFT calculations concerning the adsorption behavior and
stability of the Co-precursor on the {\sio} surface under different
hydroxylation conditions.

\begin{figure}
\begin{center}
\sglcolfigure{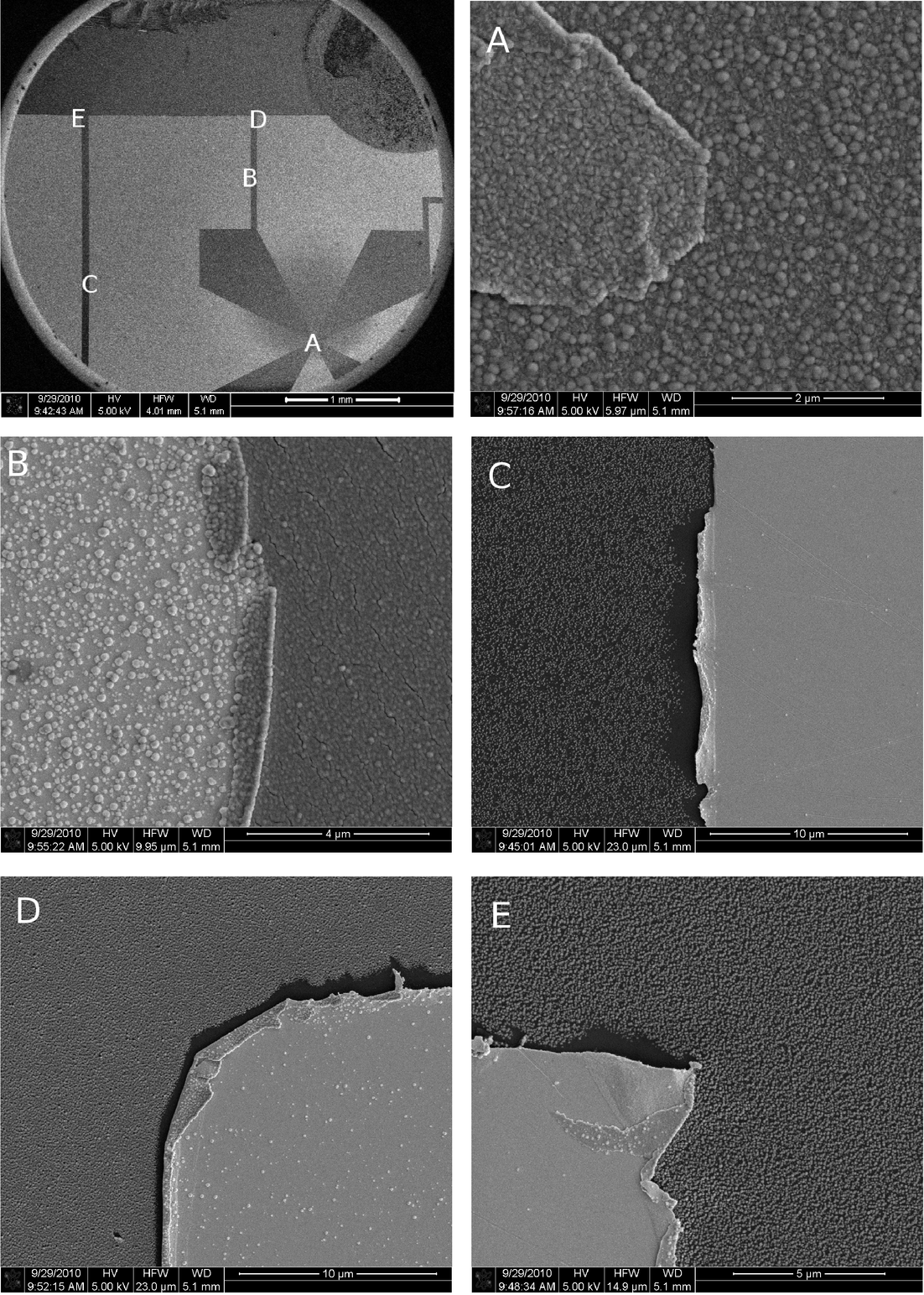}
\caption{SEM images of Co deposited on the plasma pre-treated silicon
oxide and gold. The picture on the top left is
an overview of a \sio\ surface pre-patterned with Cr/Au contact structures.
The labelling A-E indicates regions of different precursor flux which was centered at A.
The gas injection capillary is visible on the upper right.
Gold surfaces appear as bright regions, \sio\ surfaces as dark regions.
Selected area SEM images are represented in images A - E.}\label{positions}
\end{center}
\end{figure}

In a follow-up experiment we analyzed the influence of a metallic
surface, as provided by Cr/Au(20 nm/80 nm) contact structures, on this
spontaneous dissociation process (see Fig.~\ref{positions}). Inspection
of the surface on various positions of the {\sio} surface and the
Au/Cr contact structures and after 30 minutes plasma treatment and 10
minutes precursor flux exposure reveals clear differences.  In regions
of maximum precursor flux (see position A in Fig.~\ref{positions}) we
observe  slight differences in the morphology of the formed Co
clusters on the electrodes as compared to the growth on the {\sio}
surface.  In particular, a reduced average Co grain size and grain
density on the Au electrodes is observed. In regions of low precursor
flux, only small islands of the dissociation product are visible
on the Au contacts, whereas the {\sio} surface is mostly covered
(see region D and E in Fig.~\ref{positions}).
Evidently, the surface state of the plasma pre-treated {\sio} surface
provides a stronger driving force for the spontaneous precursor dissociation.

\begin{figure}
\begin{center}
\sglcolfigure{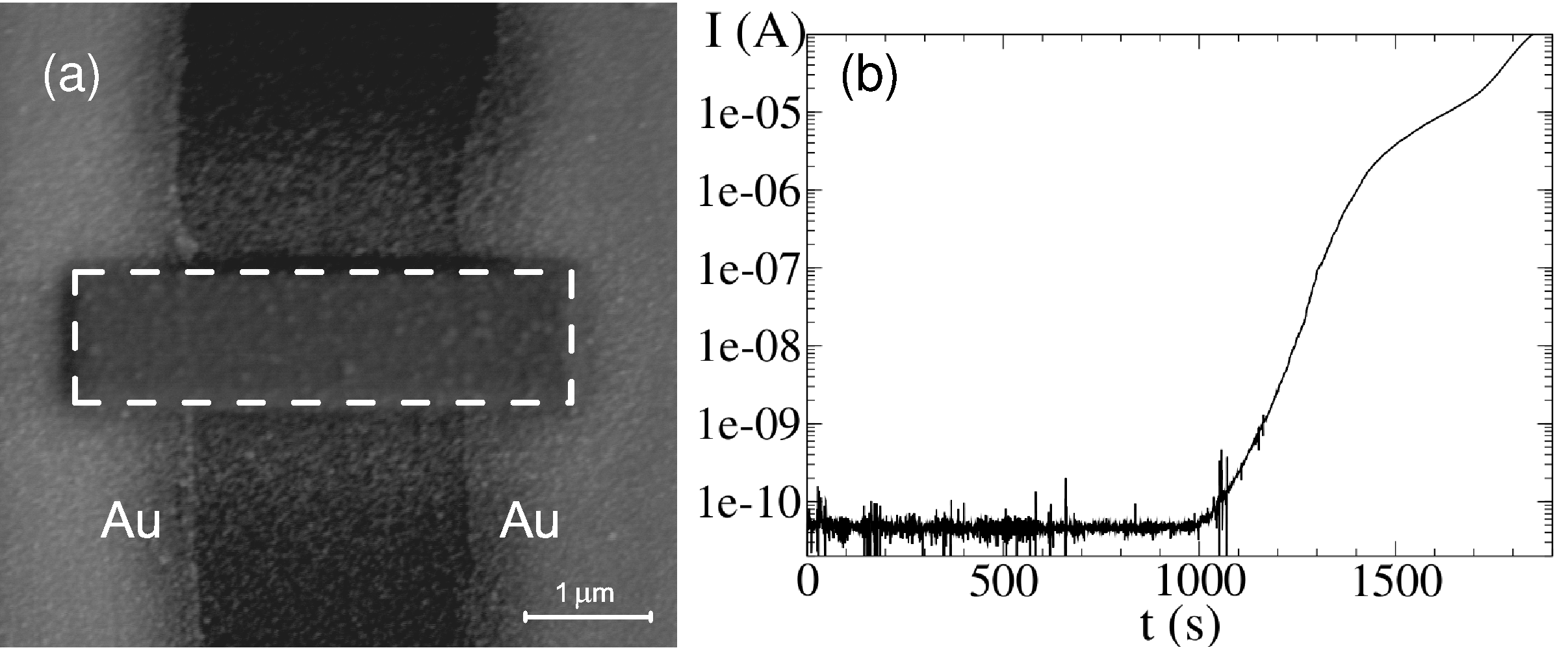}
\caption{(a) SEM micrograph of Co deposit formed after electron
  pre-irradiation of the rectangular area depicted by the dashed
  contour.  (b) Time-dependence of current flow between Au electrodes
  at fixed bias voltage (10 mV) as the Co deposit forms
  spontaneously. The current increase after closing the gas injector's valve (1200 s) 
  indicates that residual precursor molecules in the SEM vacuum chamber are continuously 
  dissociated resulting in a further increase of the thickness of the Co layer. 
  After exposing the sample to air the layer thickness was determined by atomic 
  force microscopy and found to be approximately 50 nm.}\label{SEM}
\end{center}
\end{figure}

We now turn to the results obtained on the {\sio} surface with
selected regions being pre-treated by electron irradiation.  In
Fig.~\ref{SEM}(a) we show the SEM micrograph of a Co-containing
deposit obtained in a region where the electron beam was rastered over
a rectangular area of the {\sio} surface for 10 minutes before
admission of the precursor for 20 minutes. As is evident from the
figure, a deposit between the Au electrodes has formed whose outline
represents a slightly blurry replica of the previously activated
region.
According to our Monte Carlo simulations using CASINO V2.42 \cite{casino1}
the extend of the blurred region corresponds roughly to the range of backscattered electrons.
Additional islands of the spontaneous dissociation products
are visible off the pre-treated region.  The density of these islands
drops off to zero over a length scale of about 1 $\mu$m. 

\begin{figure*}
\begin{center}
\dblcolfigure{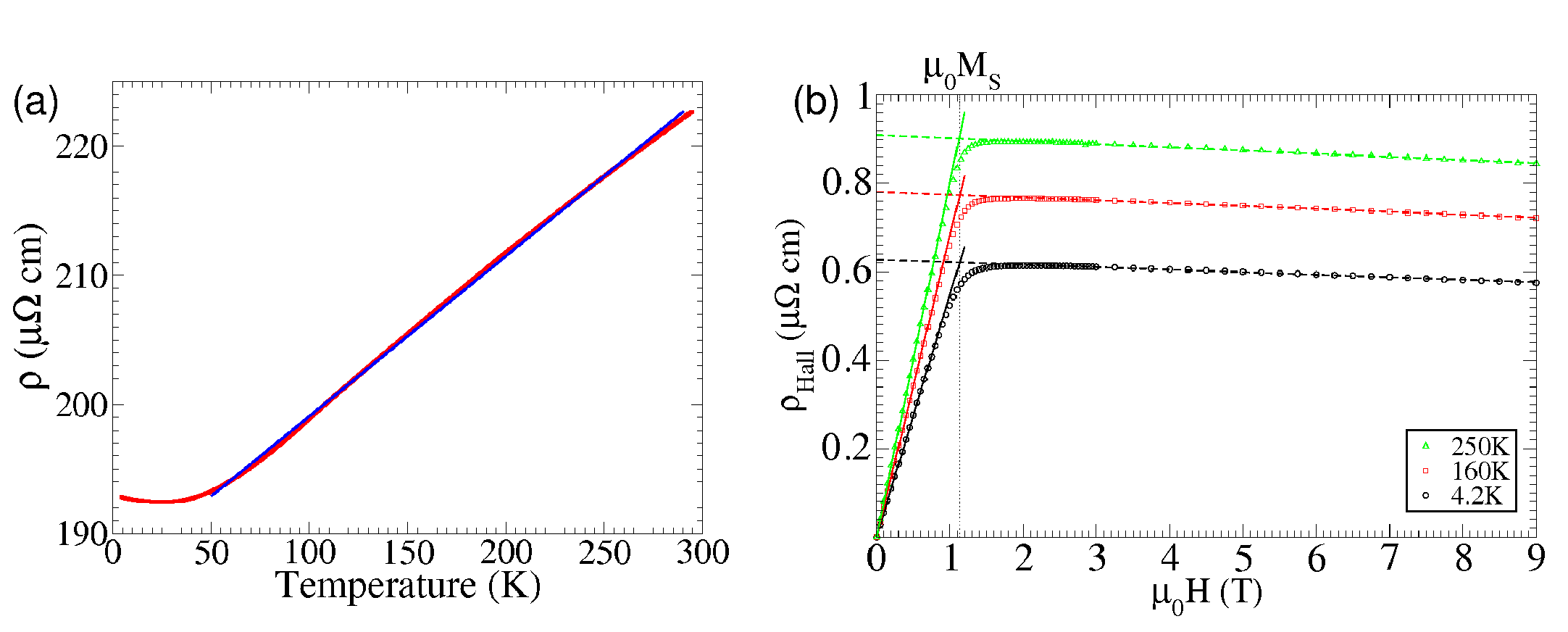}
\caption{(a)Temperature dependence of resistivity of Co deposit grown on the plasma activated \sio\
surfaces. The lateral shape of the deposit for resistivity and Hall effect measurements 
was defined by a lift-off procedure of a photolithographically defined resist pattern on which the plasma 
activated growth had been performed. The deposit height was determined to 55 nm by atomic force microscopy.
Blue line: linear fit between 50 and 290 K. (b) Hall resistivity as function of magnetic field,
measured at different temperatures as indicated. The saturation
magnetization is denoted as $\mu$$_{o}$$M$$_{s}$.}\label{measurements}
\end{center}
\end{figure*}

An energy dispersive X-ray (EDX) analysis of the dissociation products
obtained by the plasma activation and pre-growth electron irradiation
treatment reveals a Co content of approximately 95\% and 76\%,
respectively. In subsequent resistivity measurements we found a room
temperature resistivity of 223 and 480 $\mu\Omega$cm, respectively. 
This is about a factor of 5 larger than the room temperature resistivity 
found for FEBID grown Co nanowires employing the same precursor 
\cite{fernandez2009,belova2011}. A larger degree of grain boundary scattering in the 
spontaneously formed deposit, as well as a possibly higher 
carbon content may be the cause for this enhanced resistivity.
We also performed temperature-dependent resistivity measurements
(Fig.~\ref{measurements}~(a)) as well as Hall effect measurements
(Fig.~\ref{measurements}~(b)) for the sample grown on the
plasma-activated silica.  The samples grown under pre-irradiation
conditions are unstable under thermal stress and could not be measured
below room temperature. The temperature-dependent resistivity shows a
typical metallic behavior as expected for a dirty metal. From the Hall
measurement we deduced the saturation magnetization, as indicated in
Fig.~\ref{measurements}~(b), following established procedures, as
detailed in \cite{0022-3727-44-42-425001}.

\subsection{Structure and Bonding of {\coco} on {\sio} surfaces}
\subsubsection{Structure of the {\coco} molecule}
\begin{figure}
\begin{center}
\sglcolfigure{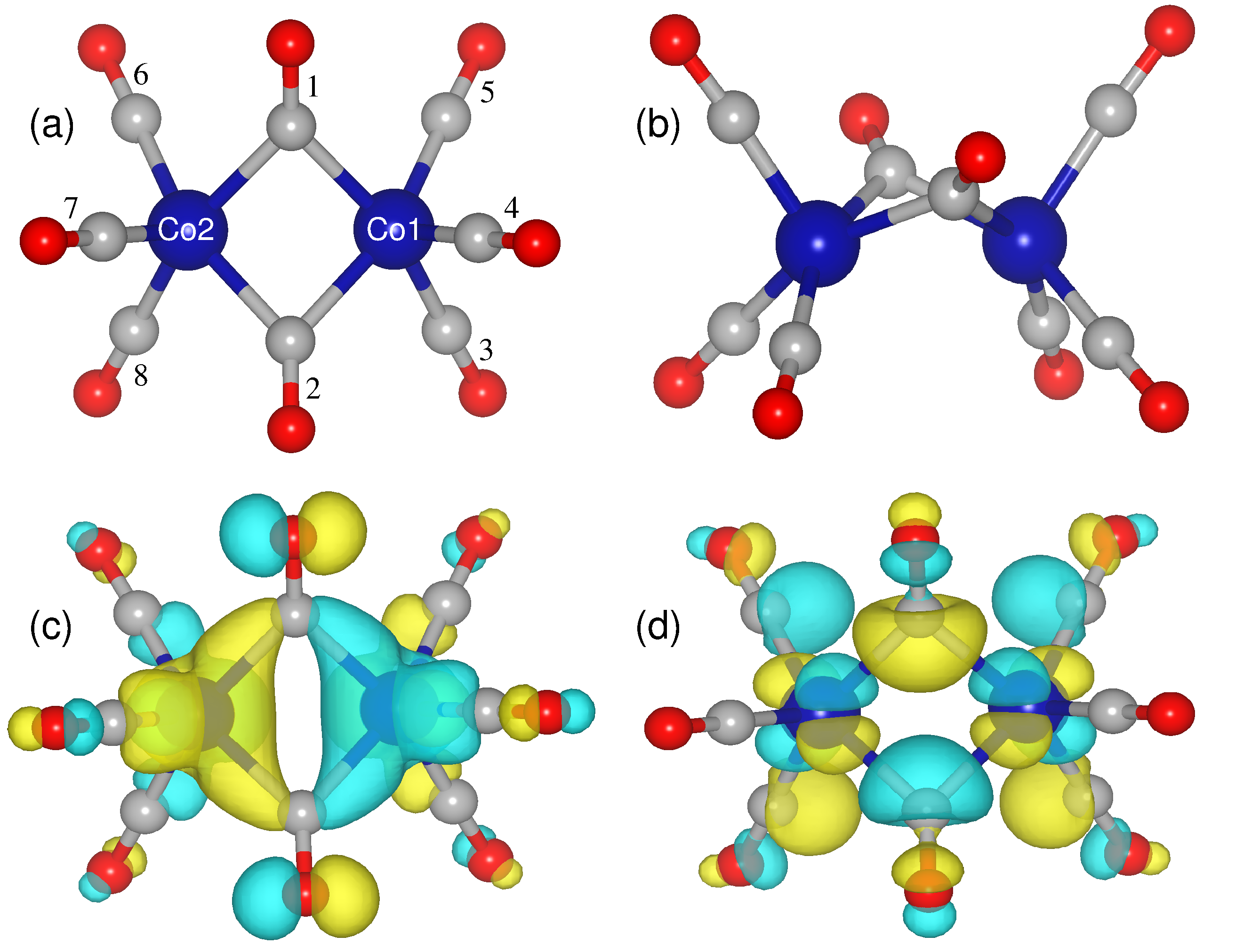}
\caption{(Color online)
(a) Top and (b) side view of DFT optimized structure of {\coco}
and its frontier orbitals (c) HOMO and (d) LUMO. Blue, red and
grey spheres represent cobalt, oxygen and  carbon atoms 
respectively.}\label{HOMO}
\end{center}
\end{figure}

The structure of {\coco} has been well studied and found to have a
distorted Fe${_2}$(CO)${_9}$ structure with one bridge carbonyl
less. Sumner {\it et al.} reported a C$_{s}$ symmetric structure
resembling the C$_{2v}$ symmetry (see Fig.~\ref{HOMO} (a)) which was
analyzed by DFT calculations \cite{EJIC:EJIC3031}. Less stable D$_{2d}$
and D$_{3d}$ isomers that do not have the bridging ligands have also
been observed in solution. \cite{G19632065,HLCA:HLCA19640470417,MK2}
The structural parameters obtained by our DFT studies, such as the
distance between the two cobalt atoms (2.52~\AA) and the distance to
the bridging (1.81~\AA) and terminal ligands (1.95~\AA) from the metal
atom, match the reported values well \cite{Sumner:a04222}. Further, we
find the D$_{3d}$ symmetric structure to be less stable by 6.9 kcal/mol
with respect to the C$_{2v}$ isomer compared to the reported value of
5.8 kcal/mol \cite{EJIC:EJIC3031}. Electronic structure analysis
indicates that the highest occupied orbital (HOMO) is dominated by Co
3$d$ orbitals (Fig.~\ref{HOMO} (b)), and the lowest unoccupied orbital
(LUMO) has a significant contribution from the 2$p$ orbitals of the
carbonyls.
\subsubsection{Bonding of {\coco} molecules on {\sio} Surfaces}
\begin{figure}
\begin{center}
\sglcolfigure{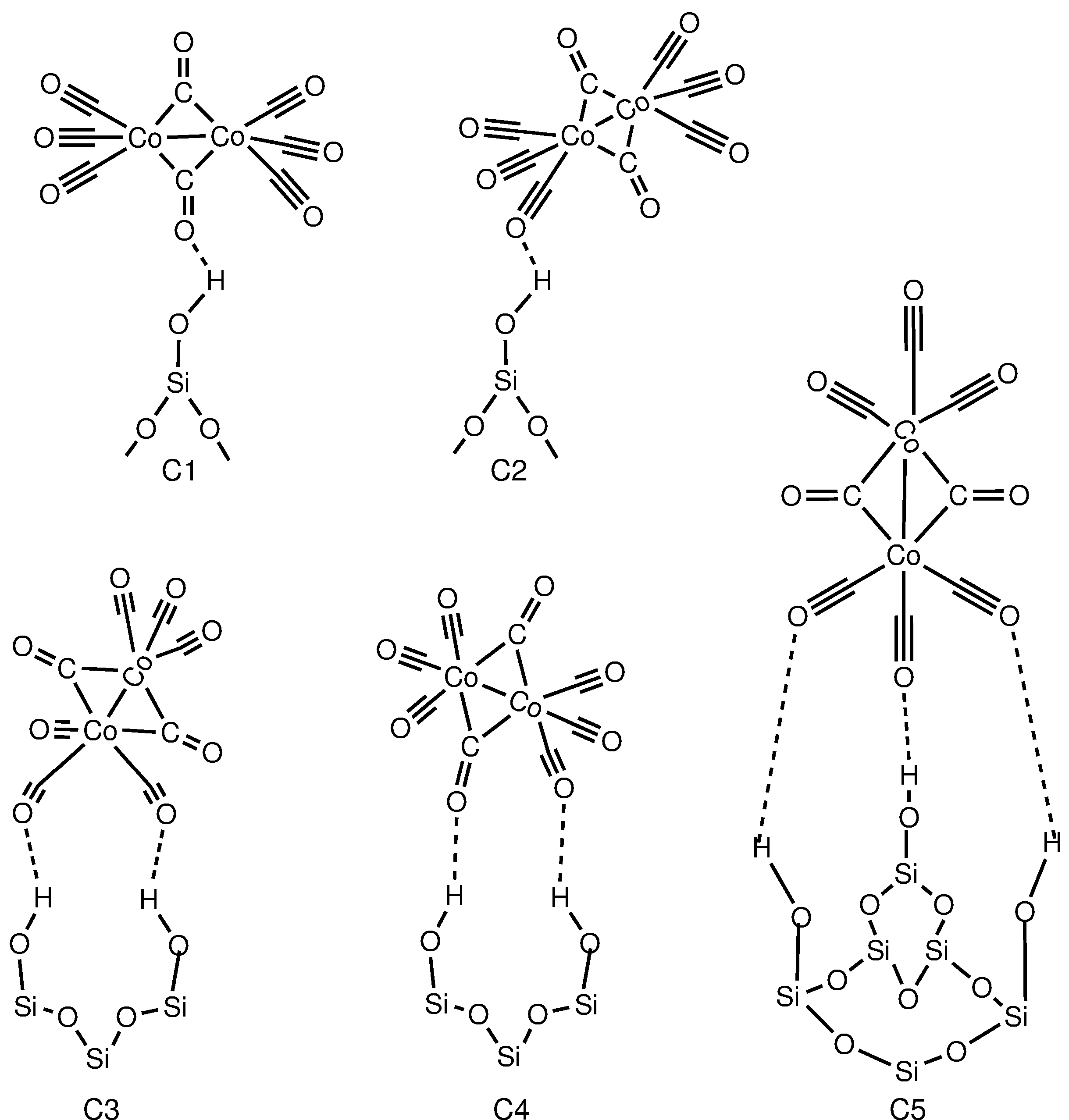}
\caption{Schematic representation of the starting configurations with
  possible {\coco} orientations considered in this study on {\FOH}
  surfaces. In {\POH} surfaces some of the OH groups are partially
  removed in order to simulate the pre-treated surfaces.}\label{configurations}
\end{center}
\end{figure}

In general, the interaction of metal carbonyls with hydroxylated
oxidic surfaces occurs through the co-ordination of the basic oxygen
of the metal carbonyls with the weakly acidic surface hydroxyls.  In
this study, we consider fully (FOH) and partial hydroxylated (POH)
{\sio} surfaces that directly represent the untreated and pre-treated
surfaces. For the {\POH} surfaces three different cases that differ in
the degree of hydroxylation corresponding to an OH vacancy concentration
of 11{\%}, 22{\%} and 33{\%} were considered depending upon the orientation
of \coco\ on the surface.~\cite{PhysRevB.84.205442}
In order to have the most stable bonding configuration of {\coco} on these {\FOH} and {\POH}
surfaces five different orientations (C1-C5 as shown in
Fig.~\ref{configurations}) were considered. These orientations take
into account the possible ways the precursor molecule can adsorb on
the surface.

The calculated adsorption energies for the different configurations of
{\coco} on {\FOH} surfaces range from -0.26~eV to -0.76~eV (see Table {\ref{tableone})
illustrating that the precursor molecule binds weakly on these surfaces.
Bonding through one of the basic bridging ligands (C1) is preferred
compared to bonding with one of the terminal ligands of the
molecule (C2). However an interesting result is obtained when
relaxations were started with the C4 configuration, where the molecule
rearranges in such a way that two of its bridging and terminal
ligands of  are oriented towards the surface (see
Fig.~\ref{FOH_POH}~(a)) with distances to the surface of
2.08~\AA$-$2.39~\AA.
The obtained distances agree well with the recently reported hydrogen
bonding distance of tungsten carbonyls with the {\sio}
substrate \cite{PhysRevB.84.205442}.
This configuration turns out to be the most stable configuration.
The difference in adsorption energy between the C4 configuration and the rest
of the configurations ranges between 0.3$-$0.5~eV. These differences
might be small under typical FEBID conditions, in particular if local
beam heating has to be taken into account. In this case the molecule
is expected to possess random orientations on the fully hydroxylated
surface. For the pre-treated {\sio} surfaces a preferential precusor
orientation is expected.  It was suggested that the weak interaction
between the metal carbonyls and the surface OH groups weakens bonding
in the molecule \cite{doi:10.1080/01614949308014607}. This is not
supported by our calculations which show negligible changes in the
Co-Co and Co-CO bonds of the precursor {\coco} of the order of
0.01-0.02~\AA.

\begin{table}
\caption{Calculated adsorption energies (in eV) of {\coco} on \sio\ surfaces.
Configurations marked with $*$ change as a result of geometry optimization and are 
discusssed in text.} \label{tableone} 
\centering                                                     
\begin{tabular}{|c|c|c|}                                       
\hline\hline 						       
Configuration & {\FOH}  & {\POH}  \\ [0.9ex]                                
\hline                                                         
C1 & -0.34  & -1.69 \\                                         
C2 & -0.26  & -0.78 \\
C3 & -0.47  & -2.46$^*$ \\
C4 & -0.76$^*$  & -3.54$^*$ \\
C5 & -0.36  & -1.12 \\ [1ex]                                   
\hline  
\hline
\end{tabular}
\end{table}

In the case of {\POH} surfaces, adsorption energies are of the order of
-0.78~eV to -3.54~eV indicating that the molecule is bound strongly
to these surfaces.
The least stable configuration is C2, where one of the terminal
ligands is bonded to the surface Si atoms. The most stable case with
an adsorption energy of -3.54~eV is obtained when relaxations were
started with C4, where one bridging and one terminal ligand were
involved in bonding to the surface.  The most interesting observation
in this case, is that the {\coco} dissociates spontaneously
into two {\cofour} molecules during geometry optimization (see
Fig.~\ref{FOH_POH}~(b)).  This
dissociation has also been observed when the molecule interacts with
the {\POH} surface with two terminal ligands (C3) and hasn't been
observed when the molecule binds either with one bridging or
terminal oxygen (C1, C2).  Although one may expect a fragmentation
of a Co-C bond to be similar to the W-C bond breaking in \wc
~\cite{PhysRevB.84.205442}, the dissociation of {\coco}
occurs by breaking the Co-Co bonds. We will discuss this process
in the next section.

\begin{figure}
\begin{center}
\sglcolfigure{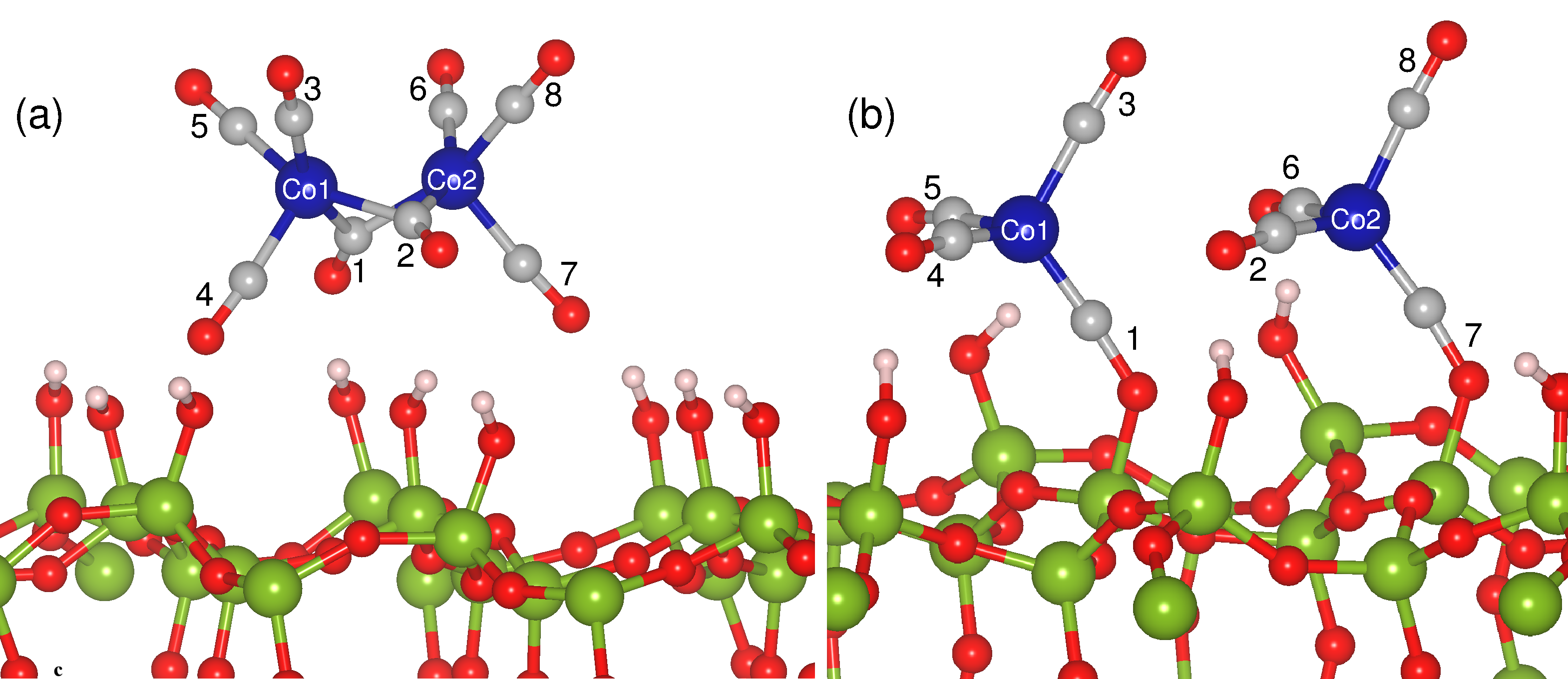}
\caption{(Color online)(a) Most stable structure of {\coco} on the (a)
{\FOH} and (b) {\POH} surfaces.  The molecule dissociates on the POH
surfaces into two {\cofour} ions bonding to a terminal Si of the surface. Green, blue, red and grey
spheres  represent silicon, cobalt, oxygen and carbon atoms 
respectively.}\label{FOH_POH}
\end{center}
\end{figure}

The above results are in agreement with our experimental observations
that the precursor molecules dissociate much more easily on the
pre-treated surfaces, as discussed in the previous section.  In
earlier experiments it was found that the decomposition of {\coco}
depends on the different number of surface hydroxyls on the {\sio}
substrates. \cite{Suvanto199925,LT199371} Although our calculations
confirm that the molecule behaves differently on {\FOH} and {\POH} surfaces,
we would like to note that the dissociation also depends on the orientation of the molecules.
For example, on the {\POH} surface the dissociation is observed only in two cases, when {\coco} is
oriented in such a way that it bonds through one terminal and bridging
ligands and when it is bonded through two terminal ligands.
In particular, we didn't observe any dissociation in C1,
which has been believed to be the prominent mode of interaction
with the weakly acidic hydroxilated surfaces in previous studies
\cite{F19888402195,doi:10.1080/01614949308014607}.
However, our results have been obtained by relaxing the
initially prepared configurations to $T=0$ directly; further
studies on the thermal stability of {\coco} on {\POH}
in C1, C2, and C5 configurations are required. 
Further, the calculated charge density for the highest occupied valence band of {\coco}
adsorbed on {\FOH} and {\POH} confirms that the
molecule retains its character on {\FOH} (compare Fig.~\ref{HOMO}~(c) and Fig.~\ref{density}~(a)),
but is strongly altered on the {\POH} surfaces (compare Fig.~\ref{HOMO}~(c) and Fig.~\ref{density}~(b)).

\begin{figure}
\begin{center}
\sglcolfigure{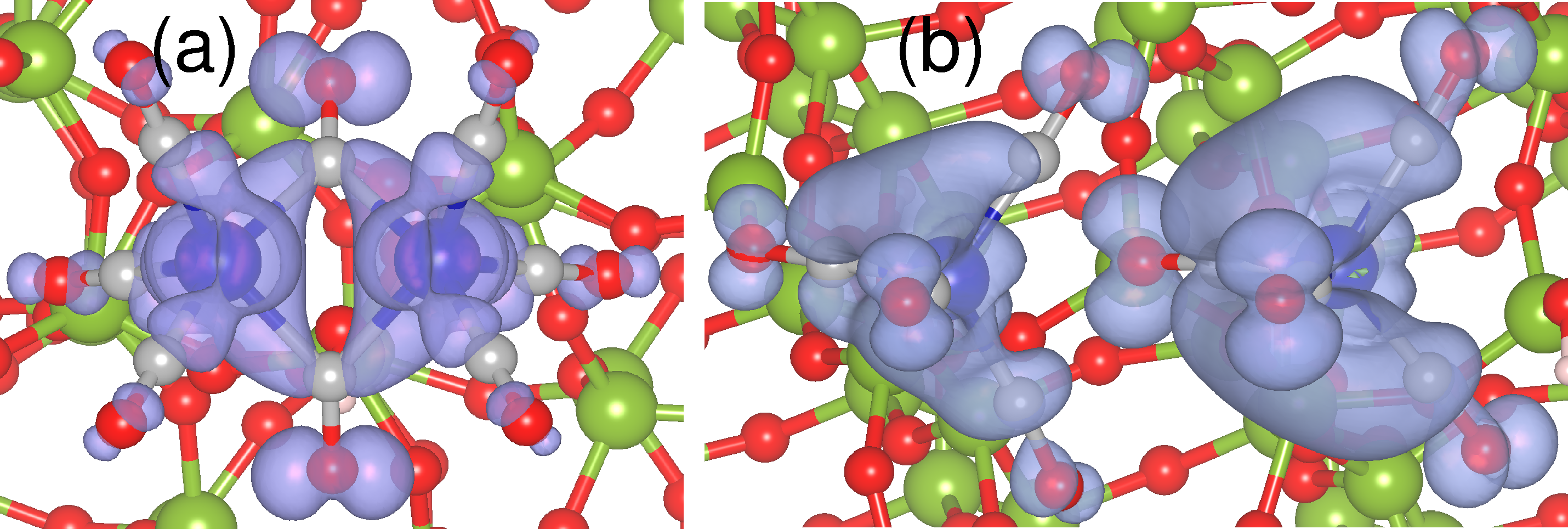}
\caption{(Color online) Band decomposed charge density for the valence band
maximum for {\coco} on the (a) {\FOH} and (b) {\POH} surfaces.}\label{density}
\end{center}
\end{figure}

\subsection{Discussion on  the dissociation and autocatalytic deposition of {\coco} precursor on {\sio} surface}

In view of the results presented in the previous section,
we will discuss here the possible reasons for dissociation and 
autocatalytic deposition of {\coco} molecules on {\sio} surfaces.

The bridging CO ligands of {\coco} possess in the free molecule
relatively higher electron density compared to the terminal ligands
(see Table~\ref{tabletwo} second column) and therefore 
are expected to be the ligands that
preferentially interact  with the dehydroxylated Si sites 
on the {\POH} surface. Our results illustrate that while
the adsorption through the bridging ligands is essential,  also 
the terminal ligands are involved in bonding to both {\FOH} and {\POH} surfaces.

Let us focus now on the dissociation process of {\coco}  on the {\POH} surface,
resulting in the formation of \cofour\ subcarbonyl motifs.
The interaction between the CO ligands of the molecule precursor 
and the dehydroxylated Si sites of the surface alters
the electronic distribution on the precursor molecule
as well as its  geometry. The changes in the electronic distribution 
are verified by the computed Bader charges on the CO ligands
(compare in Table~\ref{tabletwo} second and fourth columns)
as well as on  the Co atoms where the charge changes
from +0.74 electrons in the free molecule to +0.54 electrons upon 
adsorption. This electronic change is accompanied
by a structural change.
The bond between C and O in the bridging CO ligand weakens 
(it elongates from 1.16 \AA\ in the free molecule
to 1.25 \AA\ in the adsorbate) and  
the Co$-$C bond strengthens (it shortens from 1.95 \AA\
in the free molecule to 1.66 \AA\ in the adsorbate).
Further, the bond angle (Co-C$=$O) in the bridging ligands 
change from 140$^\circ$ to 174$^\circ$. 
In addition, the surface Si atoms acquire a more positive character
(the charge increases from +2.35 electrons to +3.2 electrons) illustrating that this
transfer of nearly one electron each from the two terminal Si sites
on to the {\coco} molecule plays a crucial role in the  
fragmentation process.
This accumulation of additional electron density on the individual
Co atoms should weaken the bonding between the two Co atoms in the precursor.
These effects such as the strong bond (Si-CO) formation followed by the 
electronic redistribution in the precursor molecule is further assisted by 
the interaction of the terminal carbonyl (see C4 in Fig.{~\ref{configurations}) 
to the surface sites that cleaves the molecules into two \cofour\ fragments.  

\begin{table}
\caption{Calculated Bader charges for {\coco} in units
of electrons 
in the  gas phase
and for the adsorbate on {\sio} surfaces. The numbers
in parenthesis identify the CO ligand as shown
in Figs.~\ref{HOMO}~(a) and Fig.~\ref{FOH_POH}. Values indicated by $*$ corresponds to the total
charge of the {\cofour} fragments}\label{tabletwo} 
\centering                                                     
\begin{tabular}{|c|c|c|c|}                                       
\hline\hline 						       
Case & Gas-Phase & {\FOH} & {\POH}  \\ [0.9ex]                                
\hline                                                         
CO(1) & -0.29  & -0.24 & -0.78 \\                                         
CO(2) & -0.29  & -0.26 & -0.24 \\
CO(3) & -0.14  & -0.09 & -0.16 \\
CO(4) & -0.14  & -0.11 & -0.22 \\
CO(5) & -0.15  & -0.06 & -0.21 \\
CO(6) & -0.15  & -0.10 & -0.23 \\                                         
CO(7) & -0.15  & -0.12 & -0.76 \\
CO(8) & -0.16  & -0.10 & -0.15 \\   
Co1   & +0.74  & +0.55 & +0.54 \\
Co2   & +0.74  & +0.55 & +0.54 \\     
Total & +0.01  & +0.02 & (-0.83/-0.84)$^*$ \\[1ex]
\hline  
\hline
\end{tabular}
\end{table}

In contrast, {\coco} binds weakly on the {\FOH} surface 
compared to {\POH} (see Table~\ref{tableone}) and retains most of its character 
similar to the free molecule (compare Fig.~\ref{HOMO} (a) and Fig ~\ref{density} (a)). 
Analysis of the charges on the CO ligands (compare in Table~\ref{tabletwo} second and third columns) 
confirm this observation.
Nevertheless, the formation of hydrogen bonds with surface
hydroxyls leads to some charge redistribution within the 
adsorbed molecule, resulting in a reduction of positive charge 
from +0.74 to +0.55 on Co. 
Also, we find minimal differences in structural parameters (of the order of
0.01 \AA).  

The above observations illustrate the fact that the weak interaction between molecule
and surface will not cause dissociation of the precursor. 
However, we would like to note that we have observed spontaneous dissociation of {\coco}
in our experiments after  extended exposure of the precursor flux (30 minutes or more).
The spontaneous dissociation under long-time exposure is likely just a sign of the instability of 
the molecule which dissociates under CO release over the intermediate {\cotwelve} at 52$^{\circ}$C. 
At lower temperature some degree of this dissociation will already be observable, in particular if there is no stabilizing CO atmosphere, 
such as is the case in a SEM vacuum chamber. 
(Moreover, the reduced neighbor coordination of the adsorbed molecules as compared to the bulk solid might speed up the dissociation process.)

In summary, our calculations confirm that {\coco} decomposes
upon its interaction with  {\POH} surfaces illustrating which might be the the first step 
occurring in this deposition process. Further \coco\ molecules possess the capability to 
deposit autocatalytically as a result of spontaneous dissociation. 
At present it is unclear how to rationalize this autocatalysis
and a detailed study based on molecular dynamic simulations is in progress 
but beyond the scope of the present work.  
We expect that the total charge on the fragmented species of {\coco}
is among the important factors that causes autocatalytic deposition.
In our calculations, these fragments possess a net charge of $-$0.84 electrons.
This charge is expected to play a similar role as the surface Si atoms on the {\POH} surface, namely
it activates the approaching molecule and triggers the autocatalytic process.
This indeed accounts for the fact that in our experimental observations 
the deposition occurs immediately on the pre-treated surface where the fragments
are formed as soon as the precursor flux is in contact with the {\POH} surface, 
and with a slight delay on the {\FOH} surface. 
However, this needs to be confirmed with theoretical simulations and 
remains as an open question that will be addressed in our future studies.

\section{CONCLUSIONS}\label{Summarize}
We report here the deposition of Co from the precursor {\coco} on two
different pre-treated {\sio} surfaces and our results 
provide an in-depth understanding of preliminary interactions
and evidence for the spontaneous dissociation.
Our observations suggest an
activation of silica surfaces which is also effective, although to a
lesser degree, on Au layers. In view of the fact that no such
spontaneous dissociation effects on Si substrates with a very thin
native oxide 
layer have been reported in previous works~\cite{0957-4484-20-47-475704,doi:10.1021/nn201517r} we are led to
assume that this surface activation process depends on both, a
modified surface termination and trapped charges. Presently it is not
clear whether the activation process observed on silica layers under
ultrahigh vacuum conditions in conjunction with the precursor {\fc}
\cite{ANIE:ANIE201001308} is also at work here.  Further,
we have also performed DFT calculations for this deposition
process considering various slabs settings and  find
that the extent of surface hydroxylation and the orientation of the 
precursor has a vital role on the dissociation and the formation of the nanocomposites. 
The so-formed
sub-carbonyl motifs during the FEBID process might be the true
precursor for the Co-rich nanocomposite formation.


\begin{acknowledgements}
  The authors would like to thanks M. U. Schmidt and M. A. Se\~naris - Rodr\'\i guez
for useful discussions and gratefully acknowledge financial support by the
  Beilstein-Institut, Frankfurt/Main, Germany, within the research
  collaboration NanoBiC as well as by the Alliance Program of the
  Helmholtz Association (HA216/EMMI).  The generous allotment of
  computer time by CSC-Frankfurt and LOEWE-CSC is gratefully
  acknowledged.
\end{acknowledgements}
\vspace{3cm}
\bibliography{Co2CO8_Article3.bib}
\bibliographystyle{bjnano}
\end{document}